\documentclass[letterpaper, 10pt, conference]{ieeeconf}

\IEEEoverridecommandlockouts

\usepackage[utf8]{inputenc}
\usepackage{graphicx}
\usepackage{subfigure}
\usepackage{multirow}
\usepackage{footnote}
\usepackage{url}
\usepackage{threeparttable}
\usepackage{placeins}
\usepackage{xcolor}
\usepackage{soul}

\title{On-Orbit Operations Simulator for Workload Measurement during Telerobotic Training}
\author{Daniel Freer, Yao Guo, Fani Deligianni, Guang-Zhong Yang}
\date{August 2019}

\author{Daniel Freer, Yao Guo, Fani Deligianni, Guang-Zhong Yang, \emph{Fellow}, \emph{IEEE}% <-this % stops a space
\thanks{*This work was supported by Engineering and Physical Sciences Research Council (EPSRC) under Grant EP/R026092/1.}% <-this % stops a space
\thanks{D. Freer, Y. Guo, F. Deligianni,  and G.-Z. Yang are with the Hamlyn Centre, South Kensington Campus, Imperial College London, London, SW7 2AZ, United Kingdom. G.-Z. Yang is also with the Institute of Medical Robotics, Shanghai Jiao Tong University, China.}
}

\begin{document}
\maketitle
\begin{abstract}
Training for telerobotic systems often makes heavy use of simulated platforms, which ensure safe operation during the learning process. Outer space is one domain in which such a simulated training platform would be useful, as On-Orbit Operations ($O^3$) can be costly, inefficient, or even dangerous if not performed properly. In this paper, we present a new telerobotic training simulator for the Canadarm2 on the International Space Station (ISS), which is able to modulate workload through the addition of confounding factors such as latency, obstacles, and time pressure. In addition, multimodal physiological data is collected from subjects as they perform a task from the simulator under these different conditions. As most current workload measures are subjective, we analyse objective measures from the simulator and EEG data that can provide a reliable measure. ANOVA of task data revealed which simulator-based performance measures could predict the presence of latency and time pressure. Furthermore, EEG classification using a Riemannian classifier and Leave-One-Subject-Out cross-validation showed promising classification performance and allowed for comparison of different channel configurations and preprocessing methods. Additionally, Riemannian distance and beta power of EEG data were investigated as potential cross-trial and continuous workload measures. 
\end{abstract}

\section{Introduction}
Telerobotic systems which are directly controlled by a human operator have been utilised in many fields, from surgery \cite{Chitwood2000RoboticSystem} to outer space \cite{Mishkin2013SpaceSpace}.
One of the largest challenges with the use of these systems is the training of novice operators. Various simulators have been created to train surgeons to operate the da Vinci system in different surgical scenarios, and have shown improvements upon traditional methods in some instances \cite{Sun2007AdvancedPlanning, Lerner2010DoesSystem}. The same is true for some common tasks in outer space \cite{Belghith2012AnTraining}, but the currently available simulators cannot modulate workload on demand in order to provide different training strategies which could translate to a broader range of tasks in outer space.\par
While many groups are investigating ways in which satellites and their robotic components can operate autonomously in outer space \cite{Rekleitis2007AutonomousSatellite}, teleoperation by humans, manually or in a shared-control mode, will still have a significant role in On-Orbit Operations ($O^3$)  and robotic control in space \cite{Zhang2017SharedDelay}. 
Currently, the most successful examples of $O^3$ come from the International Space Station (ISS), where the Canadarm2 \cite{Gibbs2004CanadaStatus} has taken much of the responsibility for building and maintenance \cite{Mishkin2013SpaceSpace}. \par
\begin{figure}
    \centering
    \includegraphics[width=.6\linewidth]{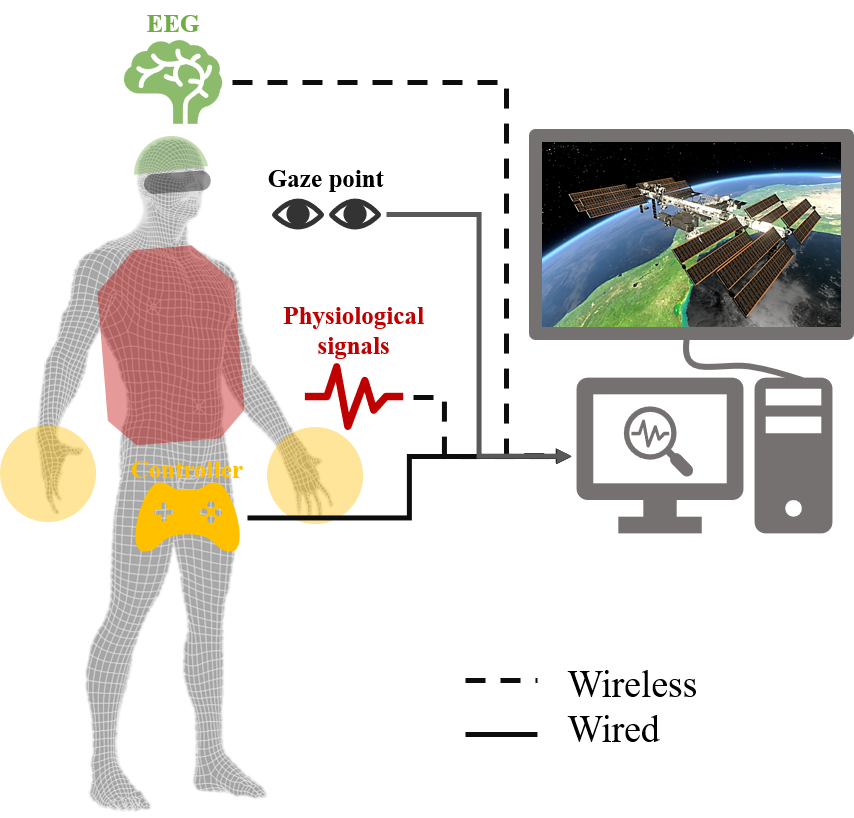}
    \caption{Demonstration of the Wearable technology for multi-modal signal monitoring of human robot interaction in $O^3$ tasks. The operator can wear this system either in space (e.g. the ISS) or at a control station on Earth.}
    \label{fig:framework}
\end{figure}
The Canadarm2 is a complex system, consisting of seven degrees of freedom (DoF), which can be considered as analogous to a human arm (Fig. \ref{fig:task}). Three DoF near the base act as the ``shoulder" before two long links with a single ``elbow" joint between them, and three final DoF that act as a ``wrist". Four controllable cameras are placed on the Canadarm2 at each end-effector and on each of the long links of the arm \cite{Mishkin2013SpaceSpace}. \par
Aboard the ISS, the Canadarm2 is controlled with the Robotic Workstation (RWS) software, which allows for multiple frames of resolution, control frames, and display frames \cite{Mishkin2013SpaceSpace}. As a result, operators of the Canadarm2 must continuously compare sensory feedback from multiple sources that may be misaligned. This can lead to increased cognitive load or degraded performance during teleoperated tasks \cite{Chintamani2011AnCues}. The visuo-spatial ability of the operator plays an important role in safe task completion, and for this reason, astronauts are required to complete hundreds of hours of intensive training with virtual simulated environments \cite{Liu2013PredictingAssessment}. \par
In this paper, we first propose a photo-realistic simulator for telerobotic training that modulates workload while simultaneously monitoring multi-modal physiological signals as shown in Fig. \ref{fig:framework}. Within this framework, we propose an improved control strategy for complex multi-camera robot manipulation. Furthermore, three realistic confounding factors are introduced to influence both workload and task performance: time-pressure, latency and obstacles. \par
Secondly, we define and evaluate the ability of objective data to measure the workload added by these confounding factors during teleoperated tasks. Measuring success and workload in complex teleoperated tasks is non-trivial, so most related studies have typically used subjective measures such as the NASA-TLX \mbox{\cite{Hart1988DevelopmentResearch}}. However, electroencephalographic (EEG) workload measurement during teleoperation has also recently been explored \mbox{\cite{Rojas2020EEGWorkloadEnvironments}}.
In this work, objective measures of operator performance are derived from the completion speed and accuracy of subtasks within the simulator, and additional measures have been defined based on Riemannian analysis of EEG data and compared with beta power.
We hope that insights from this work will eventually allow for personalised modulation of the training strategy by tweaking the confounding factors to keep users engaged, but not overstimulated or bored. \par 

\section{Canadarm2 Training Simulator}

\subsection{Canadarm2 On-Orbit Assembly Task}

\begin{figure*}[ht]
  \begin{center}
  \includegraphics[width=0.95\linewidth]{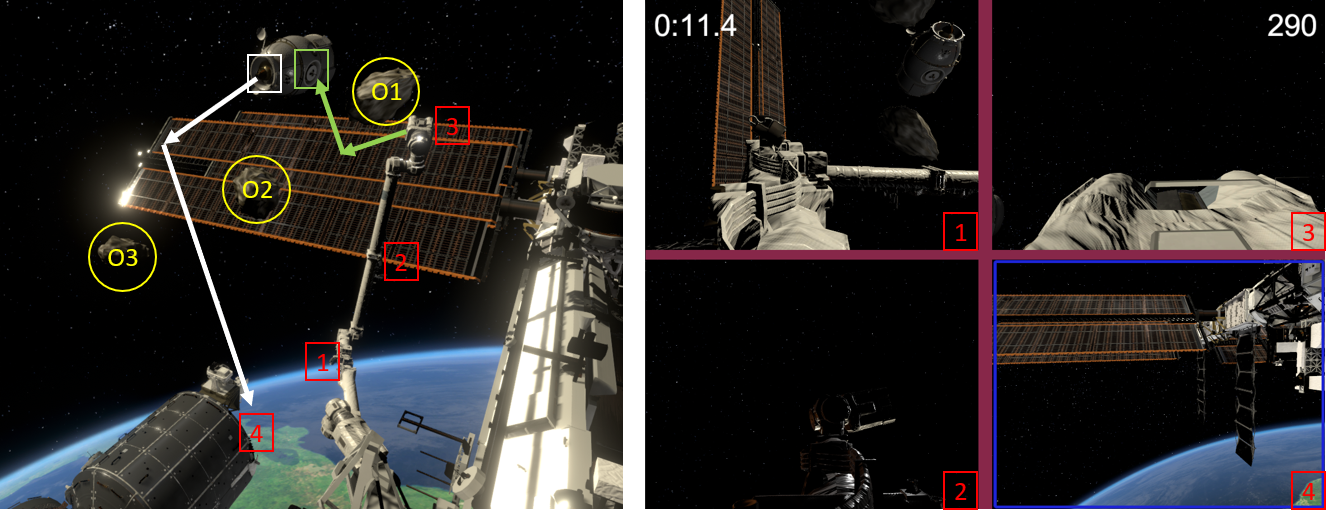} 
  \caption{The description of the task (left) and user interface (right) as shown to users before taking part in this experiment. The left part of the figure shows the two main parts of the task to be completed: 1) Navigate the end-effector to the grapple fixture on the research module and attach (green); 2) Move the research module adjacent to Columbus and dock it there (white). This picture also includes the ``debris" obstacles as a confounding factor, labelled as 'O1', 'O2', and 'O3'. The locations of four cameras are also indicated with red labelled boxes, and correspond to the similarly labelled views on the right side of the figure. The right side of the figure also shows the timer (top left corner), and scores (top right corner) for a given scenario, and the currently selected camera (camera 4) can be seen surrounded by a blue box.}
  \label{fig:task}
  \end{center}
\end{figure*}

To test the effect of confounding factors on the physiological state of telerobotic operators in training, we have developed a simulator in Unity in which users carry out the capture and addition of a new research module to the ISS. Photo-realistic 3D models of the ISS, research module and Canadarm2 were jointly developed with ZooVFX Ltd. An overview of the task can be seen in Fig. \ref{fig:task}. The task consists of several steps:
\begin{enumerate}
    \item Locate the new research module close to the ISS;
    \item Navigate the Canadarm2 end-effector to the grapple fixture on the new research module for attachment;
    \item Move the attached module to its new location adjacent to the Columbus module, while avoiding obstacles;
    \item Dock the module in its desired location.
\end{enumerate}
\subsection{Control Strategy}
In these experiments, the Canadarm2 was controlled using a standard Playstation controller, with each button and joystick commanding the actions shown in Table \ref{tab:controller}. The control methodology was based on the pose of the currently selected camera, with the left joystick movement causing up/down and left/right movement of the robot's end-effector with respect to this camera's coordinate frame. Likewise, using the L2 and R2 buttons moved the end-effector into and out of the screen for the selected camera, respectively.\par
The control strategy additionally considers joint control of three distal joints of the robot, the cameras, and the end-effector latching motion. The circle and square buttons of the Playstation controller turned the last joint of the Canadarm2 for orientation control (roll), while the right joystick similarly turned the yaw and pitch of the end-effector via direct control of the two other distal joints. The yaw and pitch of the selected camera were controlled by the D-pad. The cross button closed the latching end-effector of the Canadarm2 around the grapple fixture on the new research module, and the triangle button unlatched the end-effector and docked the module to its new desired position. \par

\begin{table}[htbp] \caption{Controller buttons mapped to robot actions}
\centering
\label{tab:controller}
\begin{tabular}{|l|l|}
\hline
\textbf{Controller Button(s)} & \textbf{Action Performed}                                                                                               \\ \hline
L1/R1                         & Select active camera                                                                                                    \\ \hline
D-pad                         & Control active camera orientation                                                                                       \\ \hline
Left Joystick                 & \begin{tabular}[c]{@{}l@{}}Move end-effector target up/down/left/right\\ in active camera coordinate frame\end{tabular} \\ \hline
L2/R2                         & \begin{tabular}[c]{@{}l@{}}Move end-effector target forward/backward\\ in active camera coordinate frame\end{tabular}   \\ \hline
Cross/Triangle                    & Connect/disconnect from grapple fixture                                                                                 \\ \hline
Square/Circle                 & Directly control final joint of robot arm                                                                               \\ \hline
Right Joystick                & Directly control 5th/6th joints of robot arm                                                                            \\ \hline
\end{tabular}
\end{table}
Users were able to cycle through which camera was selected by using the R1 and L1 buttons. The currently selected camera view was highlighted with a blue rectangular frame in the user interface. Videos from four cameras were visible to the user during the task, which corresponded to the cameras on the two long links of the robot, the end-effector, and at the docking position. Fig. \ref{fig:task} shows the user interface, including the relationship between the camera views and their positions on the Canadarm2.\par
Inverse Kinematics (IK) of the robot were achieved through iterative gradient descent using the Jacobian, which was constructed using the Denavit–Hartenberg (DH) table of the simulated Canadarm2. The robot continuously updated its joint positions until its end-effector reached within a predefined distance of its target position. 
%Because there was a limit on the angular velocity of each joint, the robot often continued moving after a movement command was finished being given. This gave the appearance of some latency, though additional latency was also added to the system, as described in Section \ref{sec: latency}. 
Only the IK of the first 5 links was considered, as any joints past this point were considered to be part of the wrist and could be independently controlled as described above. Dynamics of the robot arm and research module were not considered in this version of the simulator, as it was assumed that the Canadarm2 would be able to comfortably achieve the torques necessary to move the research module through microgravity. However, the joint velocities of the robot were limited in order to provide more realistic visual feedback for this type of teleoperated task. \par
As compared to the control strategy used aboard the ISS, our implementation automatically couples the end-effector control with the coordinate frame of the selected camera. This coupling allows the user to intuitively control the robot while watching any view of the task, rather than having to control the robot in a coordinate frame that may be misaligned from the most useful viewpoint. As a result, this control strategy should reduce mental workload because the operator will no longer have to consider multiple coordinate frames at a time while performing a given $O^3$ task. Additionally, this control strategy does not require the user to select different control modes, as control of the end-effector position, distal joints, and cameras are all included in a single controller, and can even be carried out simultaneously. The only selection the user has to make is which camera coordinate frame is active.\par
\subsection{Confounding Factors Affecting Space Teleoperation}
\subsubsection{Latency} \label{sec: latency}
While many $O^3$ tasks may be performed from a short distance away, for example by an astronaut within the ISS to the Canadarm2, some procedures may also be carried out over a long distance. Ground control often carries out the setup of a Canadarm2 task to save time for the astronauts. In these cases, ground control must cope with between 0.3 and 0.5 seconds of latency, which prevents direct teleoperation. Latency increases with distance, so any teleoperation from ground control to the moon (3-10 seconds) or Mars (20 minutes) would have more issues with latency. Expert users have reported that latency significantly affects their understanding, performance, and efficiency during teleoperated task completion, among other human factors \cite{Wojtusch2018EvaluationTeleoperation}. \par
In our implementation, we considered latency values of 0, 0.5, 1.0, and 1.5 seconds to determine whether different amounts of latency had graded effects on brain function and performance, however the simulator allows for selection of any latency value. For a given trial, the chosen latency value was constant throughout the trial. Latency was achieved by only allowing the commands from the controller to take effect once a command's timestamp was behind the processing time by more than the latency value. As a result, the listed latency values are round-trip, meaning that they are the difference in time between when the user commands robot movement and when they see the visual feedback that the robot has, in fact, moved. These latency values were chosen to approximately match the round-trip latency from Earth to the ISS, which would be 0.6-1.0 seconds if doubling the values mentioned above. While 1.5 seconds of latency may be unrealistic for this particular scenario, it may still be valuable to find how higher latencies affect mental workload during teleoperated tasks in the event that a future task needs to be carried out from a longer distance (e.g. a similar robot in a higher orbit).

\subsubsection{Obstacles}
% The presence of debris in space originating from previous missions has become a risk to current and future space missions, causing damage to spacecraft and denying access to certain orbits \cite{Liou2006RisksDebris}. While Active Debris Removal missions are crucial to preventing these downsides, consideration of how robots may be operated in the presence of debris or obstacles is additionally a valuable exercise. \par
One concern for all teleoperated tasks is the collision of the robot with objects in the scene that are either not in the field of view, or were accidentally hit by the operator. Collisions with such obstacles could result in damage to the robot or the object with which it collided. For space robots, any damage could cause mission failure and wasted resources, while for other tasks such as surgery, which may need to be teleoperated into space in the future, such collisions could result in death or injury of the patient. \par
In the case of the Canadarm2 operating outside of the ISS, the main obstacles to consider would be other parts of the ISS, but could additionally include debris that was left in space after previous missions \mbox{\cite{Liou2006RisksDebris}}, or other relatively small man-made or natural bodies that are orbiting in Low Earth Orbit. To explore this idea, three asteroid-like structures were added into the simulated environment within the workspace of the Canadarm2 to be treated as obstacles for task completion. In the task these structures were static, but would move in their proper physical direction if the robot arm collided with them. The ISS and the research module were also modelled as physical objects which could collide with the robot arm, the asteroids, or each other. Any collision of the robot arm or research module with these obstacles resulted in a 100 point decrease in the user's overall score for that trial. As a result, these obstacles provided a means to control the difficulty of the task. The timings of the initial collision with the obstacles were sent via LabStreamingLayer (LSL) and recorded for comparison between conditions and to determine whether any changes in physiological signals occurred at this time. 
% \footnote{\url{https://github.com/sccn/labstreaminglayer}}

\subsubsection{Time Pressure}
Astronauts' time in space is an invaluable resource, with longer missions being associated with serious health effects such as muscle atrophy and effects from radiation. For this reason and because some tasks may have time limitations due to power requirements or other complications, $O^3$ should be as efficient as possible, whether controlled from the ground or on orbit. Previous research in surgery has shown that adding time pressure has a large effect on the mental state of surgeons during routine laparoscopic tasks \cite{Modi2019AssociationSurgery}. As part of this study we investigate if similar mental changes are seen during this simulated task. \par
During the experiments, a timer on the screen counted up toward a time threshold of 4 minutes. The counting timer turned yellow, then red in color to inform the user that they were approaching task failure. Additionally, the user's score decreased by 10 points for every 10 seconds of task performance, encouraging speed. Whenever time pressure was added as a confounding factor, the program automatically stopped if the user had not completed the task in time. This often resulted in lower scores, as the final docking of the module increased the score for a given run.
\subsection{Performance Measures}
Throughout the task, users were scored on their performance. The task starts with the user at 300 points, which decreased 10 points for every 10 seconds the user took to complete the task, and additionally decreased by 100 points with any obstacle collisions. While time is one important factor to determine operator performance, teleoperated tasks also require precision when the robot is interacting with other objects. For this reason, at the robot's initial connection to the research module and the final docking to the Columbus module, a score related to the quality of the connection was determined and added to the users score. The quality (Q) of the connection was calculated using $Q=100/(dist+0.1) + 5000/(\theta_{dist} + 5)$, where $dist$ is the Euclidean distance between the executed and ideal position for each subtask, and $\theta_{dist}$ is the angular distance in degrees. $\theta_{dist}$ was calculated by comparing the quaternions of the ideal and actual attachment configurations in Unity. For the initial grapple of the research module, the robotic end-effector needed to come within 1 meter of the grapple fixture, while for docking 3 meters was the distance requirement. There was no angular requirement for either step. \par

\section{Data Collection}
\subsection{Experimental Protocol}
10 healthy subjects (2 females, 8 males) were asked to complete three blocks of the task while wearing an EEG cap, a PupilLabs eye-tracking system, and a physiological sensor \cite{Rosa2019AApplications} which measured heart rate, galvanic skin response, temperature and body movement. Ethical approval was received (ICREC-18IC4816), and all subjects gave informed consent. Tested subjects had varying levels of experience with 3D simulations and teleoperation, though all could be considered novices in each of these domains. While this may have affected the performance of subjects relative to one another, workload differences within a given subject should still be detectable, so for most metrics, subject normalisation was utilised. While all data was recorded from the subjects during the tasks, this paper only provides in-depth analysis of the data sensed from the EEG cap. 
Some subjects were unable to complete all three blocks due to time constraints or discomfort, so for these cases only the first two blocks were recorded. The three blocks of the task were:  
\begin{enumerate}
    \item Familiarisation Block: Users complete the task without any obstacles until they are able to complete the task in under 4 minutes. Then, obstacles are added for one run before moving onto the time pressure block.
    \item Time Pressure Block (nine total trials): Users complete the task in the presence of obstacles, once with time pressure, once with 0.5 seconds of latency, and once with neither, in a randomised order. This block was completed three times for each subject.
    \item Latency Block (six total trials): Users complete the task with obstacles and latency. The latencies considered were 0.5, 1.0, and 1.5 seconds, and were assigned in a random order for each subject. This block was completed two times for each subject, with the second time using time pressure as an additional factor.
\end{enumerate}

\begin{figure}[tbp]
    \centering
    \includegraphics[width=1\linewidth]{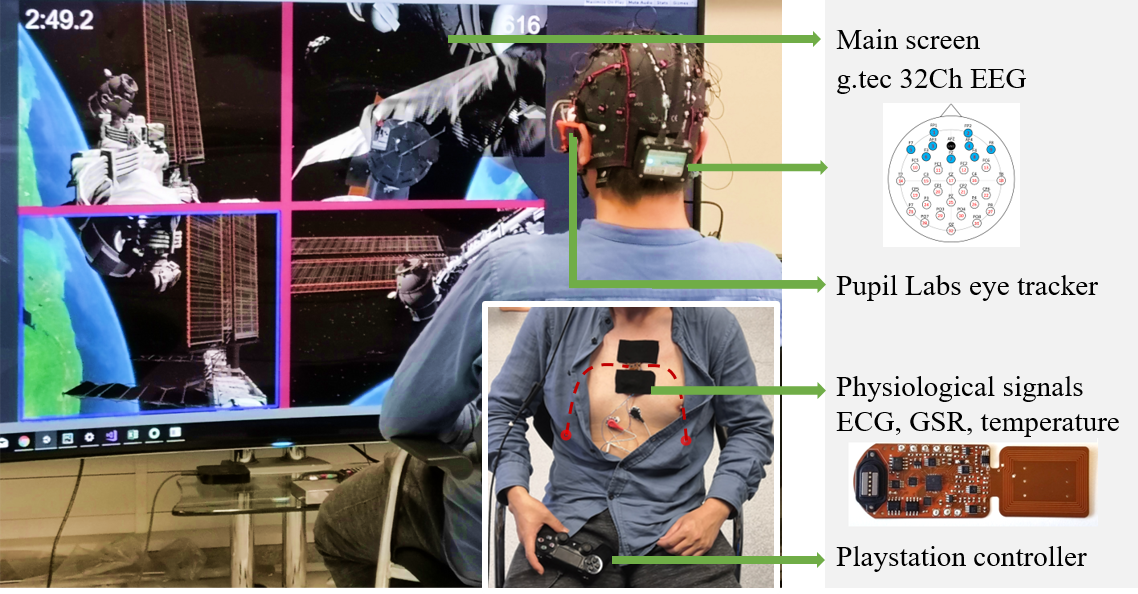}
    \caption{Experimental Scenario: The subject was asked to finish the simulated teleoperation task with a hand held controller. During task completion, physiological signals were monitored using a g.tec 32-channel EEG cap, a Pupil Labs eye tracking system, and a flexible chest patch acquiring ECG, GCR, skin temperature, and acceleration data. The multi-modal data was synchronised by LSL.}
    \label{fig:experiment}
\end{figure}

\begin{figure}
    \centering
    \includegraphics[width=0.95\linewidth]{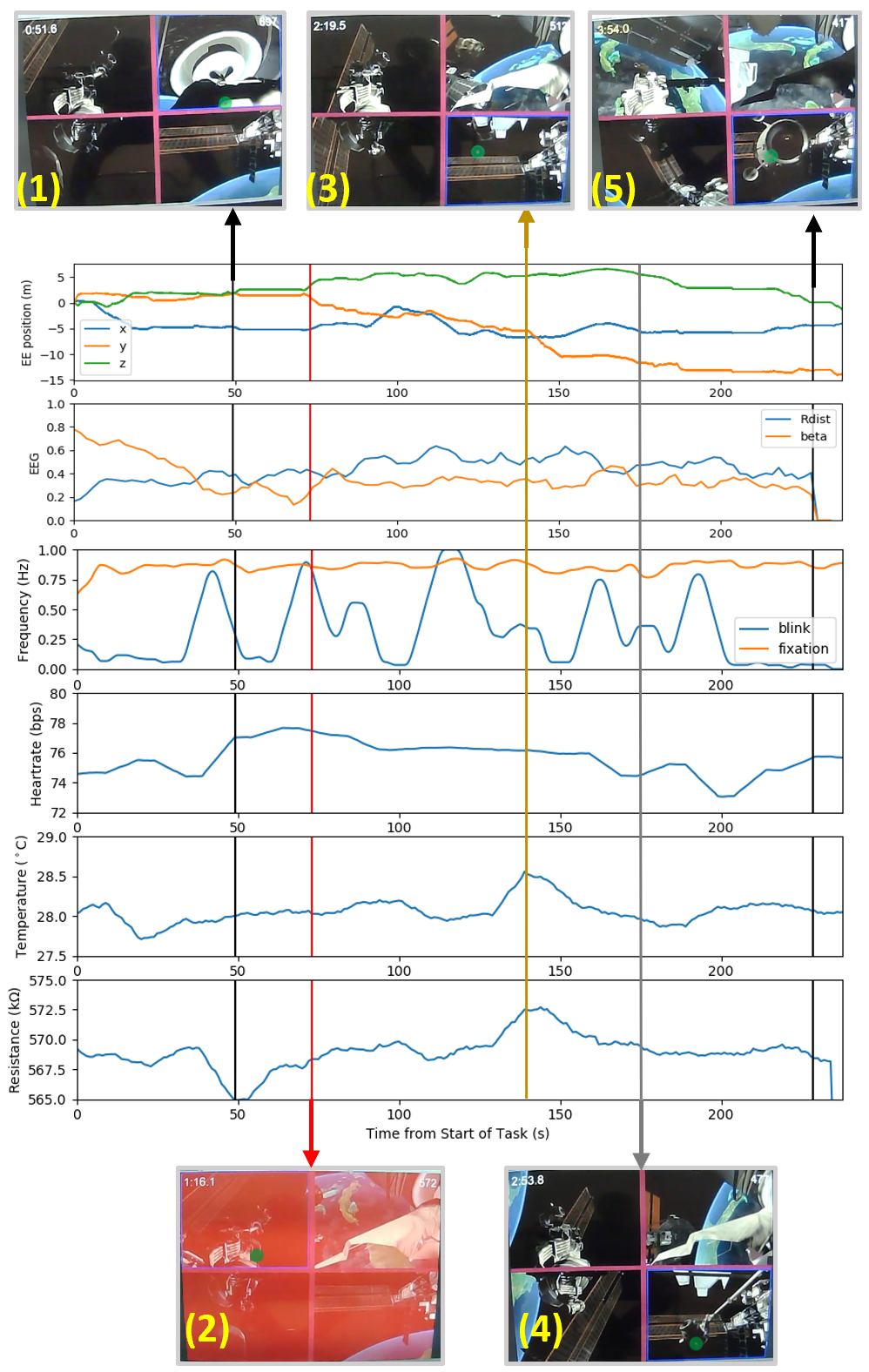}
    \caption{Multimodal sensor signals and corresponding task images from a typical run through the simulated task. The signals from top to bottom are: End-effector displacement, normalised EEG beta power and Riemannian distance from the low workload class, blink and fixation frequencies, heart rate, temperature, and skin resistance (GSR). The highlighted moments from the task include: 1) Grapple 2) Collision 3) Changing control strategy 4) Stop translation and start rotating the component 5) Several seconds before the final docking. }
    \vspace{-10pt}
    \label{fig:signals}
\end{figure}

The signal changes during a typical run through the task can be seen in Fig. \ref{fig:signals}, which notes several key points during the process, including the initial grapple, a collision, a change in control strategy, final rotation of the module, and docking to finish the task.  

\subsection{Brain Signal Collection, Processing, and Classification}
EEG signals were captured at 250 Hz through a g.tec g.Nautilus 32-channel wet EEG cap and recorded through Python using LSL. Before recording, the impedance of all channels was verified to be less than 30 k$\Omega$. All recorded EEG data were notch filtered at 50 and 100 Hz due to electronic noise which was noticed in the Fourier transform of the signals. \par
After this, three different methods of removing two known artifacts of eye blinks and muscular (EMG) artifacts were explored. The first method was Independent Component Analysis (ICA) using the MNE library \cite{GRAMFORT2014446}, considering FP1 as an EOG channel for artifacts removal. The artifacts removed from the signal by ICA preprocessing included eye-blinks and ECG, while also removing kurtosis and skewness. A second method was to bandpass filter the signal between 2 and 60 Hz using an MNE infinite impulse response (IIR) filter. A third method loosely followed the B-Alert system \cite{Berka2007EEGTasks}, which has been regarded as one of the best workload classification systems. It utilises a wavelet transform to deconstruct a signal, evaluates and amends the signal based on the deconstruction, then reconstructs it without the higher and lower frequency components. In our implementation, the signal was deconstructed around 6 frequency levels corresponding to approximately 2, 4, 8, 16, 32, and 64 Hz, then reconstructed without the highest and lowest frequency components. While the original authors additionally normalised data containing eye-blinks and rejected data that contained too much EMG noise, this was not possible for our system. Their method for rejecting eye-blinks relied on a large amount of  training data from a similar workload task with many subjects, while we did not have access to such a dataset. Also, in our task muscular movement of the fingers occurs throughout the task, so a high amount of EMG artifacts were often present. Rejecting each of these datapoints would result in too little data. \par
After preprocessing, channels were selected to reduce the dimensionality of the input signal and ensure that the covariance matrices were Symmetric and Positive Definite (SPD). The same channels were selected for all subjects for a given run and were evaluated experimentally. The different channel combinations tested are shown in Fig. \ref{fig:channel_select}, and include central diamond, central x, frontal, parietal, parallel, and rocket configurations. More channel combinations were additionally considered, but the presented configurations were the most informative in terms of determining which regions were most useful for improving classification performance. \par
Finally, classification was performed using a Riemannian Minimum Distance to the Mean (MDM) classifier. While this classifier has been used in other EEG classification tasks such as motor imagery \mbox{\cite{Freer2019AdaptiveProtocols, Barachant2012MulticlassGeometry}}, to the author's knowledge this paper is the first to explore its application to workload detection, and more specifically added workload due to latency.
Other classifiers, such as linear DFA, were also considered, but tended to fit only to a single class (the low workload class) due to class imbalance. This problem is not as prevalent in the Riemannian MDM classifier.\par
Different methods of data processing and classification were evaluated using Leave-One-Subject-Out (LOSO) cross-validation, where the data from 9 of the 10 subjects were used as training data, and the remaining subject's data was used as the test dataset. Data were considered in 2 second non-overlapping windows, and were labelled under three separate paradigms. Under the \textbf{5-class paradigm}, labels corresponded to low workload (LW), time pressure (TP), 0.5 seconds of latency (0.5s), time pressure + 0.5 seconds of latency (0.5s+TP), and trials with 1 or 1.5 seconds of latency (HighLat). Under the \textbf{latency paradigm}, only two classes were considered: those with any amount of added latency, and those without. Similarly, under the \textbf{time pressure paradigm}, the two classes considered were with and without time pressure. For all paradigms, each time window in a trial was given the same label, despite the fact that some parts of these trials may induce more workload than others. Presented F1 scores are the average of each class F1 score. \par
Derived from the Riemannian classifier, Riemannian distance to the LW class was also further analysed as a potential continuous workload measure. This value was compared to beta power, which has recently shown a positive correlation with workload during teleoperation \mbox{\cite{Rojas2020EEGWorkloadEnvironments}}. The main consideration for our method was the Riemannian distance between each 2s window of data and the average Riemannian covariance matrix of the LW class, which is denoted as $d_0$. Our hypothesis for this metric is that with higher workload, this distance should increase, as the data should become less similar to the minimum amount of workload.

\section{Results}
\subsection{Subject-specific Performance Measures}
There was high variability in performance measured across the 10 subjects. Final scores ranged from an average of 15.1 (Subject 1) to an average of 959.83 (Subject 3), while average time to task completion similarly ranged from 351 seconds (Subject 8) to 151 seconds (Subject 4).
With 123 total collisions, the most common was the research module (RM) colliding with the first obstacle (O1), which occurred 55 times. The second most common was the collision of the Canadarm2 (C2) with the body of the ISS, which occurred 29 times. Analysing the reasons for these collisions could help us to develop feedback mechanisms which could prevent them from occurring in a real-life scenario. \par

\subsection{One-way ANOVA on Performance Measures}
\label{sec-anova}
To examine how the simulator-defined performance changed with the addition of each confounding factor, we first conducted one-way Analysis of Variance (ANOVA) on each measure $x_i$ as listed in Table \mbox{\ref{tab:anova}} w.r.t. to each confounding factor $F$, where $F\in$\{\textit{Latency}, \textit{Time~Pressure}\}. As trials w/o \textit{Obstacles} only occurred in the familiarisation block, the comparison between w/ and w/o \textit{Obstacles} introduced bias, and thus it is not considered here. To reduce subject-specific differences, z-score normalisation was first performed for each subject $j$ and each metric $i$, where we have $\hat{x}_i^j = (x_i^j - \mu_x^j)/\sigma_x^j$. To perform ANOVA on a given factor $F$, the performance $x_i$ from all the samples were categorized into two groups \{w/o \textit{F}, w/ \textit{F}\}. In this case, we define all the experiments with 0.5, 1.0, and 1.5 seconds of latency as belonging to the \{w/ \textit{Latency}\} group.  As there only exist two categorical groups, the one-way ANOVA is equivalent to a standard \textit{t}-test. We repeated the experiment for each performance measure and report the $p$-value results in Table \mbox{\ref{tab:anova}}. It can be observed that the operators tended to collide with obstacles more often \{w/ \textit{Latency}\}. While the operators were under \textit{Time Pressure}, they finished each $O^3$ subtask more quickly, as can be seen by a decrease in \{Grasp time, Dock time\}.

\begin{table}[!b]
\caption{P-value for simulator-defined performance measures with and without confounding factors}
\setlength{\tabcolsep}{2pt}
\centering
\begin{tabular}{|l|p{5.1cm}|l|l|}
\hline 
\multirow{2}{*}{Category} & \multirow{2}{*}{Performance measure $x$} & \multicolumn{2}{c|}{Factors $F$} \\ \cline{3-4}
& & Latency & TP \\ \hline
\multirow{2}{*}{\scriptsize{$X_{eff}$}} & \scriptsize{Grasp time: time from the start to the grasping} & .733    & .016*$\downarrow$  \\
& \scriptsize{Dock time: time from the grasping to the docking}  & .157    & .000**$\downarrow$        \\ \hline
\multirow{4}{*}{\scriptsize{$X_{prec}$}} & \scriptsize{Grasp distance: distance error of the grasping point}   & .139    & .285          \\
& \scriptsize{Grasp angle: angle error of the grasping pose}     & .943    & .902          \\

& \scriptsize{Dock distance: distance error of the docking point}   & .675    & .741          \\
& \scriptsize{Dock angle: angle error of the docking pose} & .817    & .386          \\ \hline 
\multirow{3}{*}{\scriptsize{$X_{score}$}} & \scriptsize{Grasp score: score after grasp the research module}  & .205    & .664          \\
& \scriptsize{Dock score: score from the grasping to the docking}                  & .973    & .691          \\
& \scriptsize{Final score: overall score per trial}                     & .175    & .290      \\ \hline
\scriptsize{$X_{coll}$} & \scriptsize{No. of collisions: collisions with obstacles per trial}               & .016*$\uparrow$  & .285          \\
 \hline    
\end{tabular}
\begin{tablenotes}
\item[] \scriptsize **$p \leq .005$, *$.005 < p \leq .05$; \textit{eff} - efficiency, \textit{prec} - precision, \textit{coll} - collision
\item[] $\uparrow$, $\downarrow$ indicate the mean value changes of \{w/ \textit{F}\} compared to \{w/o \textit{F}\}
\end{tablenotes} 
\label{tab:anova}
\end{table}

\begin{figure}[htbp]
\centering
\includegraphics[width=0.7\linewidth]{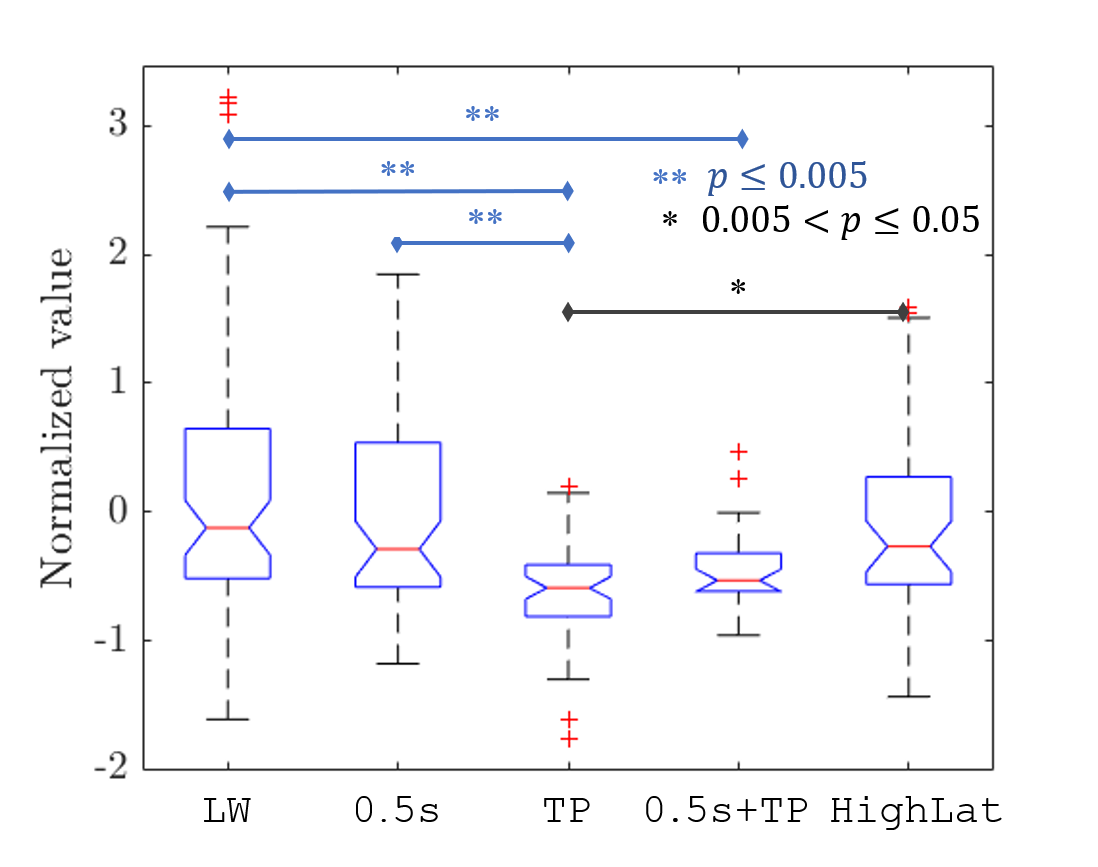}
\caption{ANOVA results for pair-wise comparison of $X_{eff}$ across 5 groups.}
    \label{fig:ANOVA5class}
\end{figure}

Next, we conducted one-way ANOVA on the performance measures with 5 workload classes, which enabled a pair-wise evaluation of the null hypothesis between any two groups. 
Here, we examined five categories of measures as summarised in the first column of Table \mbox{\ref{tab:anova}}, including $X_{eff}$ representing the operation efficiency, $X_{prec}$ indicating the precision of grasping and docking, $X_{score}$, and $X_{coll}$. Fig. \mbox{\ref{fig:ANOVA5class}} shows the ANOVA results of five groups on performance measures $X_{eff}$. Compared to the {low} workload group, the $X_{eff}$ measures under time pressure ({TP} and {0.5s+TP}) are significantly decreased, indicating that the time consumed was lower and the task was completed more quickly with added time pressure. In addition, this measure was also significantly lower for {TP} than for the latency groups (0.5s \mbox{\&} HighLat). For other measures, no statistical significance can be found between any two workload levels, which emphasizes the difficulties in discriminating 5-class workload levels from performance measures.

\begin{figure*}[t]
    \centering
    \includegraphics[width=0.9\linewidth]{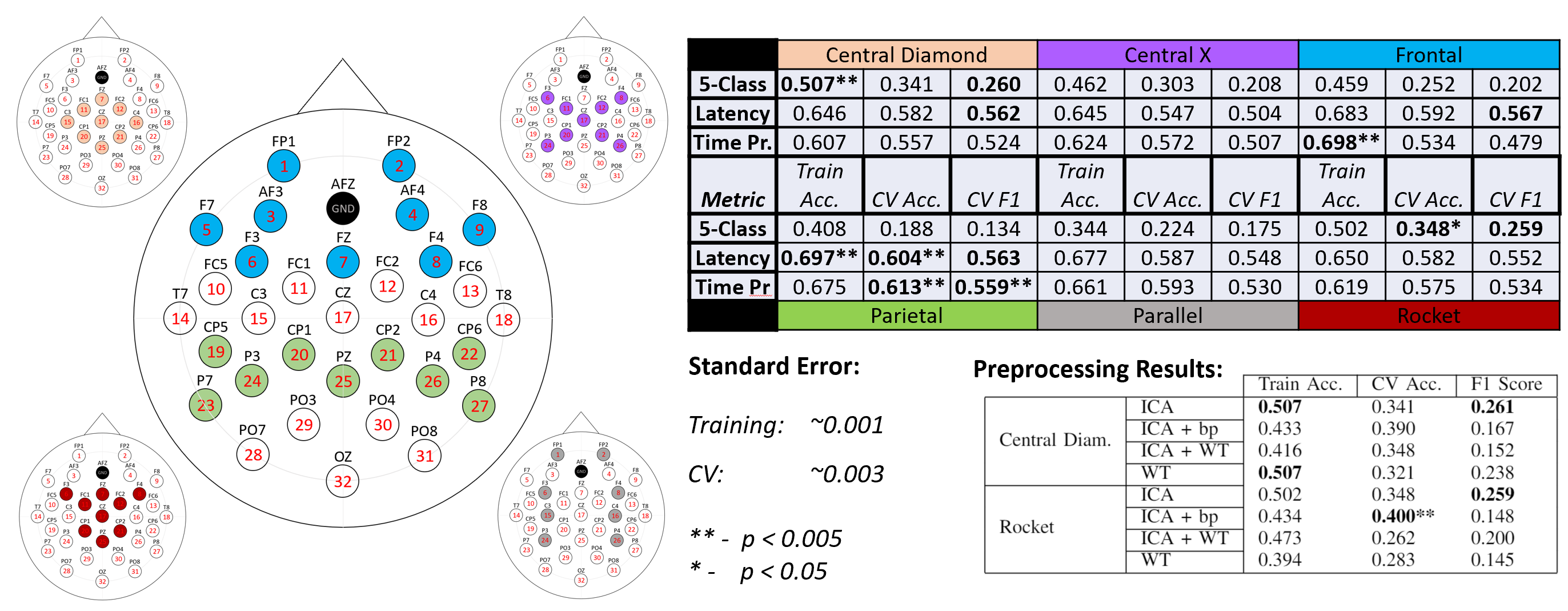}
    \caption{The layout of the 32-channel EEG cap used for recording and six different channel selection configurations considered. Classification results shown correspond to the accuracy on the training set, the LOSO cross-validation (CV) accuracy, and the LOSO CV average F1 score for a given configuration (channel selection) and paradigm (5-class, latency, or time pressure). Bold values indicate the best score for a given metric, and stars indicate statistical significance over all other conditions. The bottom right table explores different preprocessing techniques (ICA, bandpass filtering, wavelet transform, and combinations of these). All results in the top table use ICA preprocessing.}
    \label{fig:channel_select}
\end{figure*}

\subsection{Riemannian EEG Workload Classification}

Results of the channel selection process, as shown in Fig. \ref{fig:channel_select}, revealed that the most discriminative channel configurations for the 5-class problem were the central diamond and rocket configurations, which achieved cross-validation (CV) accuracies of 0.341 and 0.348, respectively, with each achieving a CV F1 score of approximately 0.26. As both of these configurations made heavy use of channels near the top and centre of the head, it seems that this region may be the most useful for determining mental workload during training for teleoperated tasks. Any configurations which did not heavily utilise this region did not achieve a CV accuracy of more than 0.252 for the 5-class problem, significantly lower than these two configurations. \par
However, when considering only the 2-class problems (with or without latency, and with or without time pressure), the most discriminative was the parietal configuration, with the best CV accuracy and F1 scores for both the latency and time pressure paradigms. Interestingly, this region showed the worst performance in the 5-class problem. The frontal region statistically tied parietal for the best CV F1 score in the latency paradigm, but showed the worst performance when discriminating trials with time pressure from those without time pressure, despite having the highest training accuracy. In all other measures, the parietal configuration was significantly better than all other configurations. \par

Using only ICA was determined to be the best preprocessing method based on the average F1 score in the 5-class paradigm. If considering only CV accuracy, including the the 2-60 Hz bandpass filter also improved performance, however this filter resulted in more predictions of the non-workload class. Because this class was more prevalent, the CV accuracy increased, but the F1 score decreased in both the central diamond and rocket configurations. Using the wavelet transform had mixed results, but showed generally poor performance when compared to only using ICA. The only exception to this was when using only the wavelet transform in the central diamond configuration. Both methods of removing higher and lower frequencies from the signal showed worse performance, indicating that features in this frequency range may be indicative of high mental workload.

All statistics were determined using an unpaired t-test between the highest value for each condition and the next highest values, using the standard error of the mean and the number of observations for each condition.

\subsection{Workload Change over Trials}
\label{subsec:TrialChange}
To further analyse the data, we investigated how workload changed with respect to the number of trials completed for a given subject. This can be viewed as measuring improvement, as subjects that have completed more trials will have more experience, and thus their mental workload is expected to be lower. Another potential hypothesis would be that as the subject performs more trials, they become fatigued, thus reducing their capacity for mental workload. In either respect, we expect to see that increased number of completed trials should correspond to lower relative workload (meaning lower beta power and Riemannian distance), while overall scores on the simulator should increase.\par
\begin{figure}[!t]
    \centering
    \includegraphics[trim={0 0 1cm 1cm},clip,width=0.7\linewidth]{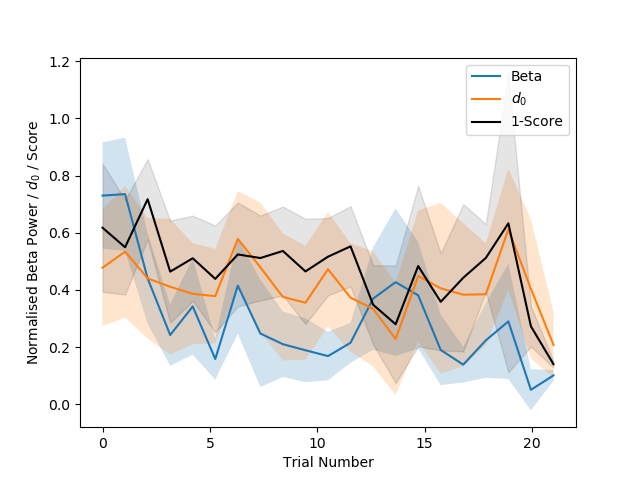}
    \caption{Subject-normalised beta power and $d_0$ plotted over number of trials completed per subject. Additionally, the value for 1 minus the normalised Score value for each trial number is plotted. The shaded areas represent the 95\% confidence intervals.}
    \label{fig:fatigue}
\end{figure}

From \mbox{Fig. \ref{fig:fatigue}}, we can see that subject-normalised beta power most notably decreases within the first few trials, indicating that there is much improved task understanding during the ``Training Block". This difference is not clear when considering normalised $d_0$. However, when comparing each metric to the simulator-defined overall score, it seemed that the $d_0$ followed the inverted performance of the operator more closely than beta power. This indicates that while $d_0$ may not measure workload as accurately as beta power, it might be better at predicting teleoperation performance. However, this needs to be more fully investigated.

\subsection{EEG as a Continuous Workload Measure}

To investigate the reason for the low classification accuracies, we performed a more thorough analysis of Riemannian distance ($d_0$) and whether it could feasibly be used as a continuous measure for workload.  In these calculations, because of the results from Section \mbox{\ref{subsec:TrialChange}}, the familiarisation block was removed in order to provide a balanced comparison with beta power. Additionally, data was z-normalised, with all data outside of each class's 95\% confidence interval removed before being processed by ANOVA.

\begin{figure}[htbp]
\centering
\begin{subfigure}[$d_0$]{
\includegraphics[width=0.49\linewidth,trim={0 0.5cm 0.5cm 0.5cm},clip]{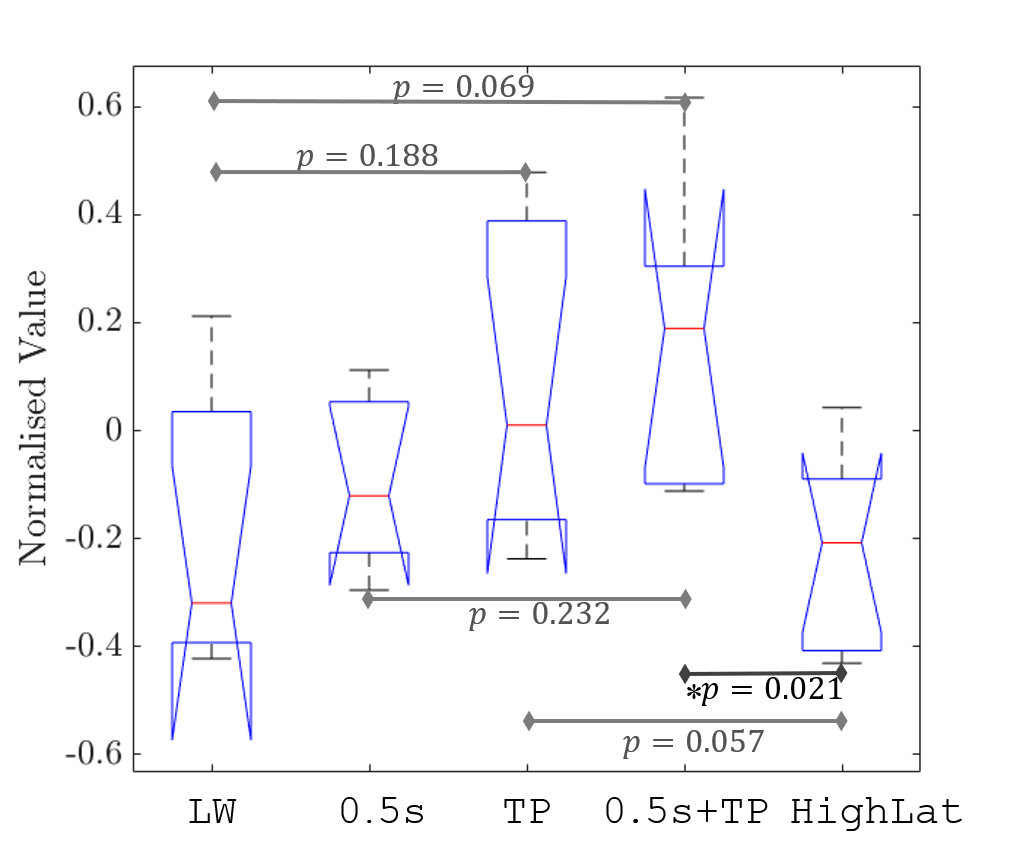}}
\end{subfigure}\hspace{-10pt}
\begin{subfigure}[Beta power]{
\includegraphics[width=0.49\linewidth,trim={0 0.5cm 0.5cm 0.5cm},clip]{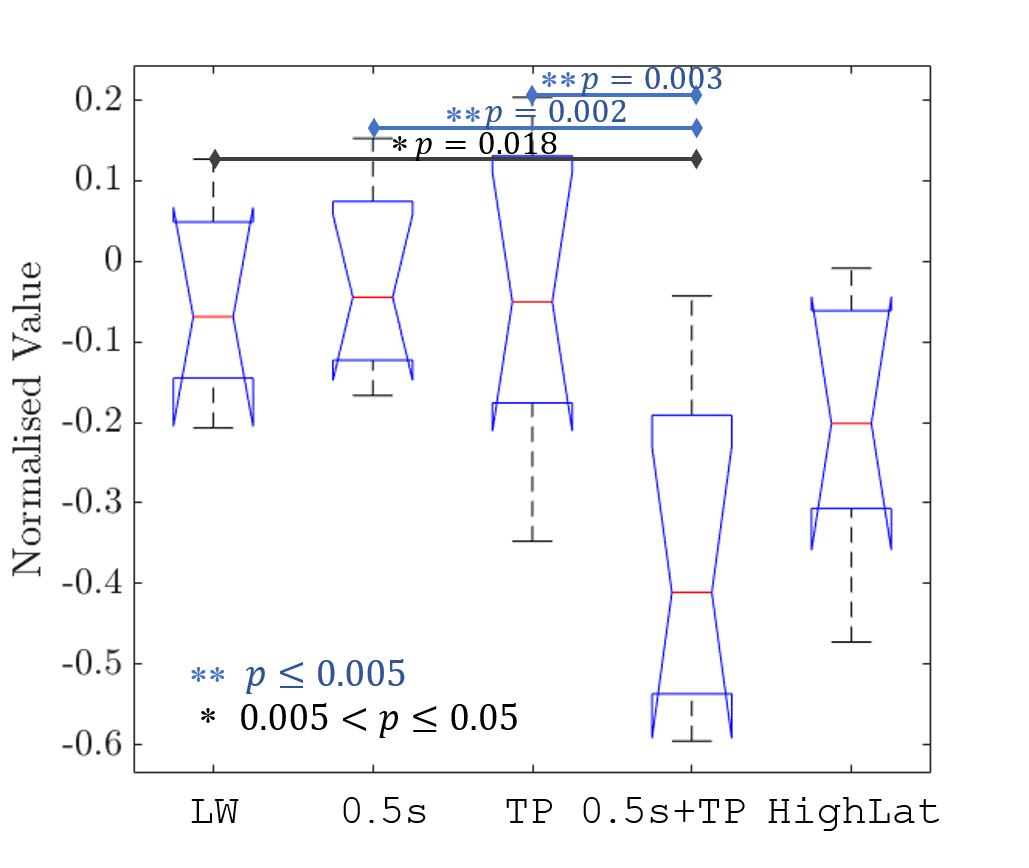}}
\end{subfigure}
% \begin{subfigure}[average dist 1]{
% \includegraphics[width=0.3\linewidth]{average_dist1_5groups.png}}
% \end{subfigure}
% \begin{subfigure}[average Distance To Trial Avg vairance]{
% \includegraphics[width=0.3\linewidth]{averageDistanceToTrialAverage_vairance.png}}
% \end{subfigure}
% \begin{subfigure}[Trial Avg To Class0 Avg]{
% \includegraphics[width=0.3\linewidth]{TrialAvgToClass0Avg.png}}
% \end{subfigure}
\caption{ANOVA results of pair-wise comparisons on the Riemannian distance ($d_0$) and beta power metrics across 5 groups.}
    \label{fig:ANOVA_eeg}
\end{figure}

The results from this investigation can be seen in Fig. \mbox{\ref{fig:ANOVA_eeg}}.
Here we can see several conditions which are significantly different when considering 2-second windows, indicating that our simulator has successfully modulated the mental workload of subjects by adding time pressure and latency. Excluding the HighLat condition, $d_0$ increased approximately as expected, with the LW class showing the lowest value and increasing with the 0.5s condition, then again with the TP and 0.5s+TP conditions. However none of these differences were statistically significant because the variance within each class was very high. The only significant difference in $d_0$ was a drop between the 0.5s+TP condition to the HighLat condition. In some subjects high latency might increase difficulty to a  point that the subject disengage from the task. In terms of beta power, however, the 0.5s+TP condition was significantly lower in value than the LW, 0.5s, and TP conditions, which were all very similar in normalised value. These results reveal that while both of these measures may indicate mental workload in some capacity, they are likely picking up on different workload features, which must be further investigated in the future.

\section{Conclusions and Future Work}
We have proposed a human-robot-interaction framework that modulates workload and allows for objective performance evaluation and multi-modal bio-signal monitoring. Intended for use during training for telerobotic On-Orbit Operations, this framework considers factors that may be important to these operators such as latency, time pressure, and obstacles.
We present which of our simulator-defined objective measures of workload are efficiently modulated by changes in each of these confounding factors, with the most notable change being a decrease in task completion time when time pressure was added. Riemannian EEG workload classification allowed us to analyse different preprocessing methods and channel selection configurations to maximise insights into how the brain may be functioning under different workload conditions. We have additionally analysed whether EEG Riemannian distance could be valuable as a continuous workload measure and compared it to beta power. This analysis revealed that while Riemannian distance from the low workload class generally increased with workload as expected, the variance between different trials is too high for it to currently provide a reliable measure. This high variance between trials may result from the complexity of the task compared to tasks that are usually used for workload measurement, and should be clarified by collecting more subject data.\par
Future work should focus on combining the information from the various sensors utilised with the framework in order to provide reliable, real-time feedback related to workload which could enhance training protocols. This framework can also be used to study how feedback and semi-autonomous robotic control can affect the workload of operators. We additionally hope to improve the simulator by implementing multiple types of common On-Orbit Operations, and add realism by conferring with astronauts and other telerobotic operators. Through these improvements, the simulator can personalise and enhance the training of telerobotic operators around, and outside of, the Earth.\par 

\bibliographystyle{IEEEtran}
\bibliography{ref}

\end{document}